\documentclass[preprint,12pt]{elsarticle}

\makeatletter
\def\ps@pprintTitle{%
 \let\@oddhead\@empty
 \let\@evenhead\@empty
 \def\@oddfoot{\centerline{\thepage}}%
 \let\@evenfoot\@oddfoot}
\makeatother

\usepackage{amssymb}

\usepackage{lineno}
\AtBeginDvi{}
\usepackage[dvipdfm,bookmarks=true,bookmarksnumbered=true,colorlinks=true,linkcolor=blue,citecolor=blue,filecolor=blue,pagecolor=blue,urlcolor=blue]{hyperref}
\usepackage[all]{hypcap}

\journal{Astroparticle Physics}

\begin{document}

\begin{frontmatter}



\title{Fast localization of coalescing binaries with a heterogeneous network of advanced gravitational wave detectors}

\author[label1]{Yoshinori Fujii}
\ead{yoshinori.fujii@nao.ac.jp}

\author[label2]{Thomas Adams}
\author[label2]{Fr$\acute{\rm e}$d$\acute{\rm e}$rique Marion}
\author[label1,label2,label3]{Raffaele Flaminio}

\address[label1]{Department of Astronomy, The University of Tokyo, Bunkyo, Tokyo 113-0033, Japan}
\address[label2]{Laboratoire d'Annecy de Physique des Particules (LAPP), Univ. Grenoble Alpes, Universit$\acute{\rm e}$ Savoie Mont Blanc, CNRS/IN2P3, F-74941 Annecy, France}
\address[label3]{National Astronomical Observatory of Japan, Osawa, Mitaka-shi, Tokyo 181-8588, Japan}

\begin{abstract}
We present the expected performance regarding fast sky localization of coalescing binaries with a
network of three gravitational wave detectors having heterogeneous sensitivities, such as the LIGO-Virgo network.
A hierarchical approach can be used in order to make an effective use of information from the least sensitive detector.
In this approach, the presence of an event seen in coincidence in the two more sensitive detectors triggers a focused search in the data of the third, less sensitive, detector with a lower signal-to-noise ratio (SNR) threshold.
We investigate the benefit for sky localization that can be expected in such an approach, using simulated data and signals. 
We find that as the sensitivity of Virgo approaches one third of the LIGO sensitivity, the accuracy and precision of the localization can be improved by about a factor 3 when Virgo data are searched with an SNR threshold around 3.5.
\end{abstract}


\end{frontmatter}


\section{Introduction}
\label{1}

The first detection of a gravitational wave (GW) signal from the coalescence of two neutron stars by the LIGO-Virgo network in coincidence with an electromagnetic transient opened the era of multi-messenger astronomy~\cite{2017ApJ...848L..12A}.
In order to expand GW astronomy with better sky localization, better sky coverage, and more precise parameter estimation~\cite{Abbott2016,0264-9381-28-12-125023}, the global network of advanced GW detectors will be extended with KAGRA~\cite{PhysRevD.88.043007,doi:10.1093/ptep/ptx180} and LIGO-India~\cite{doi:10.1142/S0218271813410101}.
In the upcoming years, it is expected that the detectors in the network will have heterogeneous sensitivities, with detectors still at an early configuration and commissioning stage being less sensitive than the more advanced detectors.
In this paper, we investigate how to make an effective use of such a network in order to localize coalescing binaries.

Compact coalescing binaries are sources of particular interest for ground-based GW detectors.
When detecting the signals from these events, we perform sky localization in a low-latency mode in order to trigger follow-up observations, especially in the electromagnetic spectrum. 
In a nutshell, low-latency search pipelines that specifically target compact binary coalescence (CBC) sources operate in the following way: they process the data with matched filtering, based on a discrete set of templates covering a broad source parameters space, and record triggers when the SNR of the filtered data exceeds some threshold~\cite{PhysRevD.85.122006}.
Triggers coincident in several detectors are identified and used to reconstruct the sky location of the source. 
Events with sufficient significance are then communicated to the astronomy community for follow-up observations. 
The LIGO-Virgo Collaboration has been performing low-latency CBC searches with several pipelines: multi-band template analysis (MBTA)~\cite{Adams:2015ulm}, GstLAL~\cite{PhysRevD.95.042001} and pyCBC Live~\cite{Nitz:2018rgo}.
Candidate events are uploaded to the GraceDb~\cite{gracedb} and processed through the Bayesian rapid localization algorithm (Bayestar)~\cite{PhysRevD.93.024013} for fast position reconstruction, which generates probability sky maps.

With three detectors of heterogeneous sensitivities in terms of typical detection range for CBC sources, a hierarchical approach can be used in order to make an effective use of data from the least sensitive detector.
In this approach, the presence of an event seen in coincidence in the two more sensitive detectors triggers a focused search in the data of the third, less sensitive, detector with a lower SNR threshold.
In this process we look for a signal in a small time window around the time of the identified coincidence and having the same source parameters.

This paper explores the benefit that can be expected from such a hierarchical approach.
Particularly this study was conducted in the framework of the MBTA pipeline, coupled to Bayestar for source localization, but the approach and results are quite general and could apply beyond the case of a specific pipeline.
Section \ref{3} describes the settings of the simulation and the figures of merit of the localization performance.
Section \ref{4} shows the simulated results.

\section{Calculation set up}
\label{3}
This study is based on simulated data previously generated in the context of~\cite{0004-637X-795-2-105} and analyzed in order to derive the sky localization performance obtained when post-processing MBTA triggers reported in~\cite{Adams:2015ulm}. 
The data set features 248 simulated signals from binary neutron star (BNS) sources, injected into simulated detector noise designed to match the expected initial performance of the Advanced LIGO detectors~\cite{0004-637X-795-2-105}.
Although the corresponding detector sensitivity, which translated into a BNS detection range of 54 Mpc, does not match the more recent performance of the LIGO detectors, this is not an issue for this study, which depends primarily on the relative sensitivities of the detectors in the network and hardly on the absolute value of the BNS range.
Therefore our results are relevant for the current and future LIGO-Virgo network.

\begin{figure}[h]
\begin{center}
\includegraphics[width=31pc]{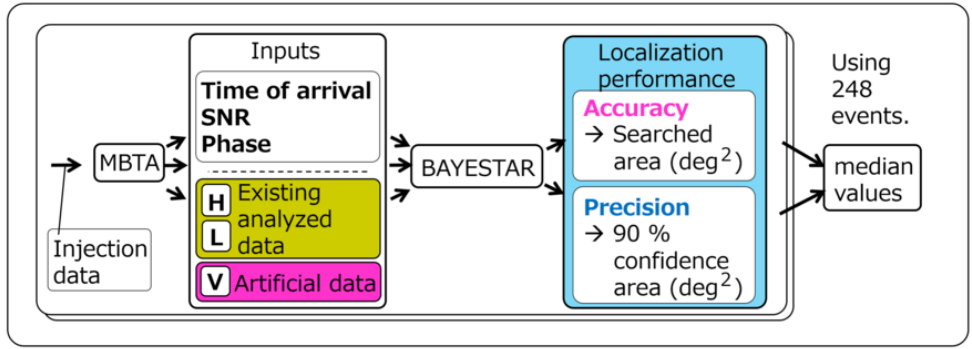}
\caption{\label{fig3_1} Overview of calculation flow. 
Using sky maps generated for 248 events processed by MBTA and Bayestar, we investigate the accuracy and precision of the localization, quantified by the median values of the searched area and 90\% confidence area respectively. 
We repeat this calculation with different sensitivities and SNR threshold values for the least sensitive Virgo detector.}
\end{center}
\end{figure}

In order to build a sky map for the location of a source, Bayestar processes information about the signal as it was detected in each detector (SNR, time and phase of arrival) and also takes as input the sensitivity curve of each detector, represented by the noise power spectral density (PSD). 
This study investigates the localization performance for the set of injections introduced above with the three-detector LIGO-Virgo network. 
While for the LIGO detectors, the trigger information (SNR, time, and phase) is extracted from running the MBTA pipeline on the data, as was done in~\cite{Adams:2015ulm}, for Virgo we use a different procedure. 
Since we want to explore several relative sensitivities of Virgo compared to LIGO, and several possibilities for the SNR threshold used in the Virgo analysis, but want to avoid running the analysis multiple times, the Virgo triggers are artificially generated in a way that emulates the result of running the MBTA pipeline and is described below.
The overview of main calculation flow is summarized in Figure~\ref{fig3_1}.

\subsection{Simulated data set and injections}
\label{3_1}
\begin{figure}[h]
\begin{center}
\includegraphics[width=31pc]{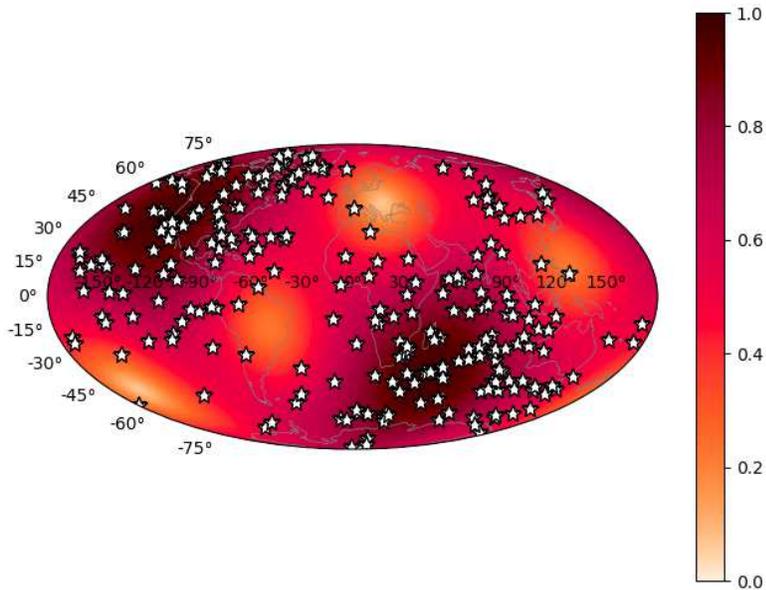}
\caption{\label{fig3_4} The sky locations of the 248 injections considered in this work are shown as stars on top of the color-coded combined antenna pattern of the two LIGO detectors.}
\end{center}
\end{figure}

The simulated data set used in this study was generated assuming noise curves based on early expectations of the 2015 performance of the detectors~\cite{0004-637X-795-2-105}.
Since we want to explore various cases for the sensitivity of Virgo, the Virgo PSD was rescaled to correspond to various relative values of the BNS range with respect to LIGO.
The 248 BNS injections had component masses uniformly distributed between 1.2$M_{\odot}$ and 1.6$M_{\odot}$, and dimensionless component spins of up to 0.05. 
Although the original population of injections composed of 10,000 events was isotropically distributed in space, we consider 248 injections which are detected in the two LIGO detectors, which biases the source directions toward directions favorable for the LIGO detectors.
The distribution of the 248 injections is summarized in Figure \ref{fig3_4}. 

\subsection{Generating artificial triggers for Virgo}
\label{3_2}
We start from 248 injections detected as HL double coincidences. 
Depending on the outcome of the targeted search in the Virgo data obtained by looking for a trigger occurring close in time and with the same parameters as the HL coincidence, each of these can either remain a double coincidence if no trigger is found in Virgo (HL case), or else become a triple coincidence (HLV case).
The latter case can appear in either one of two possibilities: The trigger found in Virgo is actually related to the injected signal (${\rm V}_{\rm i}$ case) or is related to detector noise (${\rm V}_{\rm n}$ case).
The first step of the procedure is therefore to construct a set of injections with appropriate fractions of HL, ${\rm V}_{\rm i}$ and ${\rm V}_{\rm n}$ cases.

\begin{figure}[t]
\begin{minipage}{18pc}
\includegraphics[width=18pc]{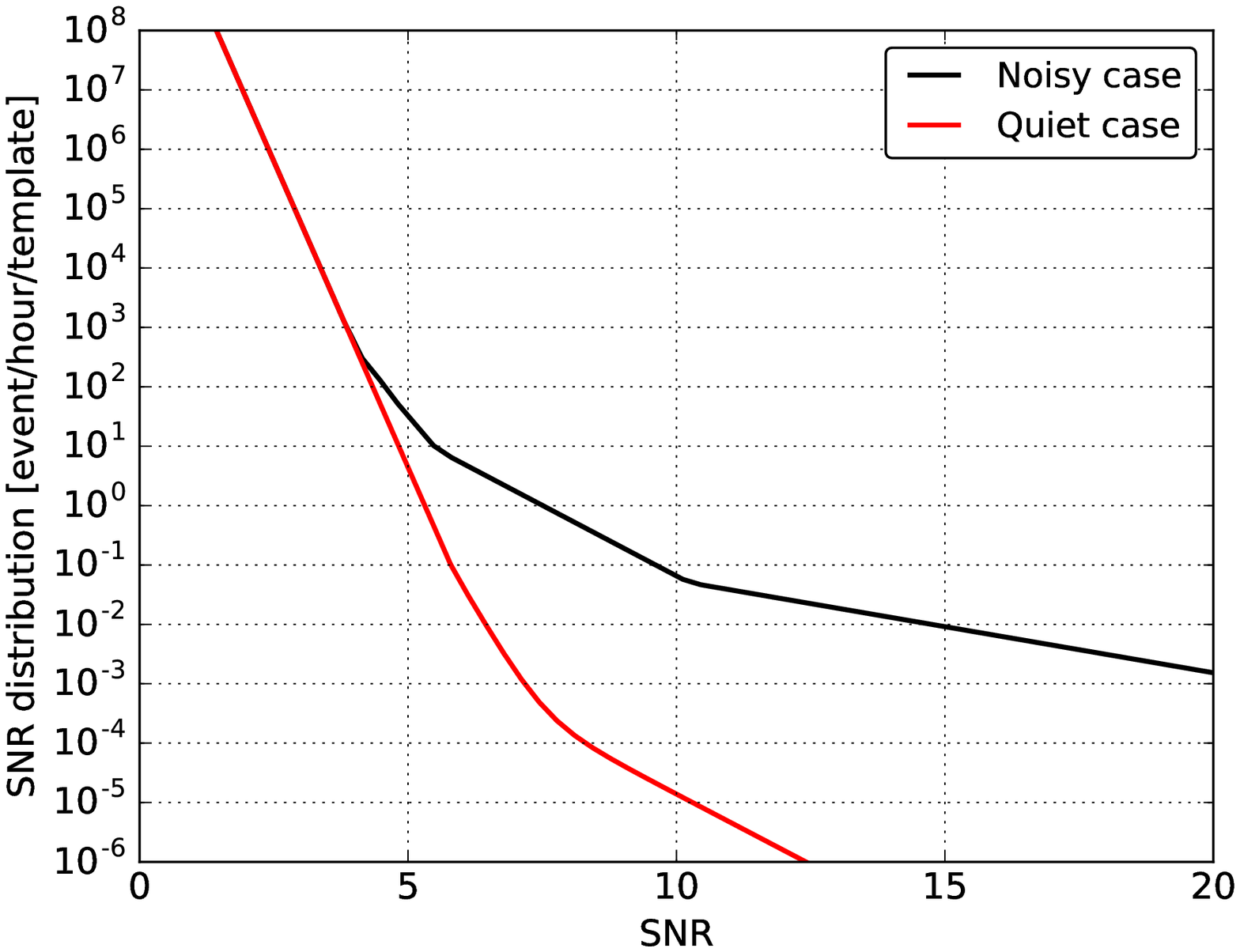}
\end{minipage}\hspace{-1.5pc}
\begin{minipage}{18pc}
\includegraphics[width=18pc]{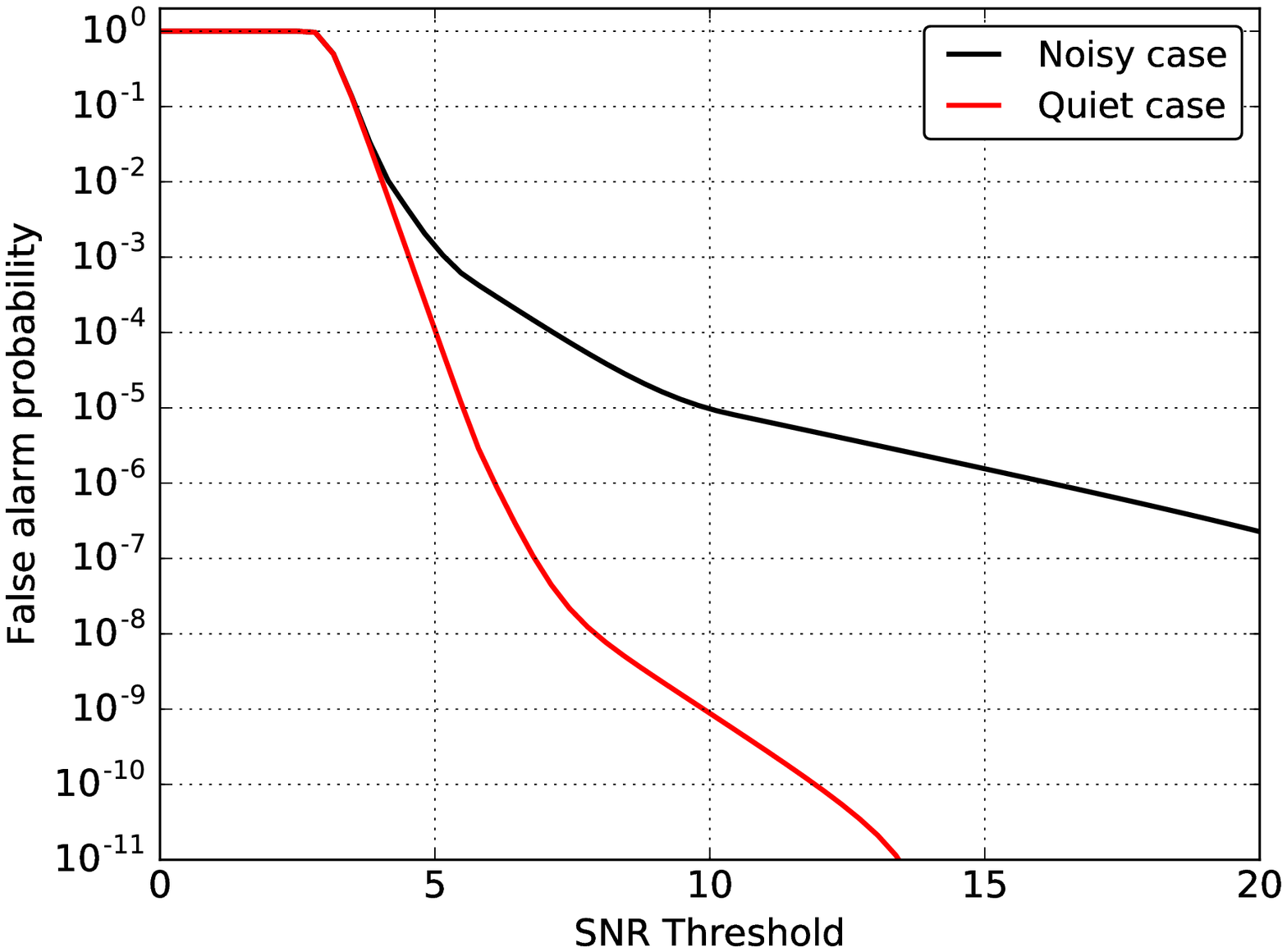}
\end{minipage}
\caption{\label{fig3_2_1} ($Left$) SNR distribution of noise triggers per hour and per template based on a measurement done during O1 and extrapolated for SNR below 6.
The red curve was obtained on quiet data and the black curve on data with excess noise. 
($Right$) False alarm probability (FAP) as a function of the SNR threshold, computed as $ {\rm FAP} = 1 - \exp(-R\ T)$, with $R$ the rate of triggers above threshold per template, derived from the distribution on the left, and $T$ = 70 ms.
At low SNR threshold, FAP saturates at about 1 since the rate R becomes quite large.}
\end{figure}

To assess the probability to get a ${\rm V}_{\rm n}$ trigger, we need an estimate of the false alarm probability (FAP) in Virgo above a given SNR threshold, for a single template and a time window of 70 ms since we consider Virgo triggers within $\pm$ 35 ms of the LIGO triggers. 
This is derived from the SNR distribution obtained by running MBTA on representative subsets of O1 data, with about $2 \times 10^5$ templates, then assuming that the trigger rate is uniform across templates, and extrapolated below the SNR threshold of 6 applied in these analyses.
The extrapolation used a Gaussian function, known to be a good approximation for the distribution of triggers at low SNR, which was confirmed by running small-scale analyses with lower SNR thresholds.
We use two data sets, one corresponding to Virgo showing nominal behavior (quiet case) and one corresponding to a time of excess noise (noisy case). 
The SNR distributions are shown in Figure \ref{fig3_2_1}, along with the FAP as a function of SNR threshold that is derived from them.

For each injection, we estimate the SNR expected in Virgo ${SNR}_{\rm \ V}^{\rm \ expected}$ from the known effective distance $D_{\rm eff}$, allowing for some statistical uncertainty:
\begin{eqnarray}
{SNR}^{\rm \ expected}_{\rm \ V}\ =\ 2.26\times {\rm (detection\ range)} \times 8/ D_{\rm eff} + {\rm Gauss}(0, 1),
\label{eq3_5_1}
\end{eqnarray}
where Gauss($\mu$, $\sigma$) is a random number derived from a Gaussian distribution with mean $\mu$ and standard deviation $\sigma$.
In this formula, the factor 8 is the SNR threshold used to define the horizon distance for an optimally located and oriented source, and the factor 2.26 connects horizon distance to detection range by averaging on location and orientation.
The ${\rm V}_{\rm i}$ case applies if ${SNR}_{\rm \ V}^{\rm \ expected}$ is above the SNR threshold in Virgo ${SNR}_{\rm \ V}^{\rm \ th}$ and there is no louder noise trigger, i.e. a random number $p_{\rm V}$ drawn from a uniform distribution between 0 and 1 is smaller than FAP.
The probability of the ${\rm V}_{\rm n}$ case depends on the probability of getting a noise trigger above ${SNR}_{\rm \ V}^{\rm \ expected}$ or above ${SNR}_{\rm \ V}^{\rm \ th}$ if ${SNR}_{\rm \ V}^{\rm \ expected}$ is smaller than ${SNR}_{\rm \ V}^{\rm \ th} $, i.e. FAP($max({SNR}_{\rm \ V}^{\rm \ th}, {SNR}_{\rm \ V}^{\rm \ expected})$). 
The procedure is summarized in Table~\ref{table_3_2_1}.

\begin{table}[h]
\begin{center}
\caption{\label{table_3_2_1}Procedure for generating coincident events. 
$p_{\rm V}$ is a random number from a uniform distribution between 0 and 1. 
${\rm FAP}_{\rm V}$ is the false alarm probability at a given SNR threshold in Virgo.
}
\vspace{1pc}
\begin{tabular}{l|l}
 \hline\hline
Conditions & Generated coincidences \\\hline\hline
if $p_{\rm V} < {\rm FAP}_{\rm V}(\ max({SNR}_{\rm \ V}^{\rm \ th},\ {SNR}_{\rm \ V}^{\rm \ expected})\ )$ & ${\rm H\ L}\ {\rm V}_{\rm n}$ \\\hline
\multicolumn{2}{l}{else} \\\hline
\ \ if ${SNR}_{\rm \ V}^{\rm \ expected} > {SNR}_{\rm \ V}^{\rm \ th}$ & ${\rm H\ L}\ {\rm V}_{\rm i}$ \\\hline
\ \ if ${SNR}_{\rm \ V}^{\rm \ expected} < {SNR}_{\rm \ V}^{\rm \ th}$ & ${\rm H\ L}$ \\\hline\hline 
\end{tabular}   
\end{center}
\end{table}

\subsection{Attributing parameters to Virgo triggers}
\label{3_3}
To produce sky maps for triple coincidences, we need to attribute parameters (SNR, time and phase of arrival) to the ${\rm V}_{\rm i}$ or ${\rm V}_{\rm n}$ triggers that supplement the HL double coincidence. 
This is done according to the procedure summarized in table~\ref{table_3_3_1}. 
For ${\rm V}_{\rm i}$ triggers we use parameters known from the injection metadata, and add ad-hoc statistical uncertainties.
For the latter we started from educated guesses based on our experience of running the MBTA pipeline, which were then slightly adjusted to get consistent sky localization performances for HL${\rm V}_{\rm i}$ cases, i.e. making sure that the fraction of injections found within the area at a given confidence level matches that confidence level. 

\begin{table}[h]
\begin{center}
\caption{\label{table_3_3_1}Procedure for attributing parameters to Virgo triggers. 
$t^{\rm measured}_{\rm LIGO}$ and ${\phi}^{\rm measured}_{\rm LIGO}$ represent, respectively, the measured time and phase of arrival at either the LIGO-Hanford or the LIGO-Livingston detector. 
For these parameters, we use the ones whose SNR is closer to the expected SNR of Virgo detector.
$\Delta t^{\rm injection}$ and $\Delta {\phi}^{\rm injection}$ describe the simulated LIGO-Virgo differences of time and phase respectively.
Random[$a$ : $b$] describes a random number uniformly drawn between $a$ and $b$.
We use 0.11 msec and 0.35 rad as typical measurement uncertainties at an SNR of 6 for the time and phase of arrival.
These values have been adjusted so that the localization areas at a given confidence level are statistically self-consistent.
}

\begin{tabular}{l|l}
\multicolumn{2}{c}{} \\\hline\hline
\multicolumn{2}{c}{${\rm V}_{\rm n}$ : Virgo trigger from noise} \\\hline
SNR & randomly drawn from the distribution shown in figure~\ref{fig3_2_1} \\\hline
Time & $t^{\rm measured}_{\rm LIGO}\ +\ $ Random[$-35$ msec : $35$ msec] \\\hline
Phase & Random [0 : 2$\pi$] \\\hline\hline
\multicolumn{2}{c}{} \\\hline\hline
\multicolumn{2}{c}{${\rm V}_{\rm i}$ : Virgo trigger from injections} \\\hline
SNR & $2.26\times {\rm (detection\ range)} \times 8/ D_{\rm eff} + {\rm Gauss}(0, 1)$ \\\hline
Time & $t^{\rm measured}_{\rm LIGO}\ +\ \Delta t^{\rm injection}\ +\ $ Gauss(0, $0.11\ {\rm msec} \times 6/{\rm SNR}^{\rm expected}$) \\\hline
Phase & ${\phi}^{\rm measured}_{\rm LIGO}\ +\ \Delta {\phi}^{\rm injection}\ +\ $ Gauss(0, $0.35\ {\rm rad} \times 6/{\rm SNR}^{\rm expected}$) \\\hline\hline
\end{tabular}
\end{center}
\end{table}

\subsection{Figures of merit}
\label{3_4}
We use the 90\% confidence area and the so-called searched area as figures of merit for the performance of the sky localization. 
The searched area is the area of the highest confidence region around the pixel of maximum probability, that includes the sky location of the injected GW signal. 
The 90\% confidence area gives the precision, whereas the searched area quantifies the accuracy of the sky localization.

\section{Sky localization with hierarchical search by 3 detectors}
\label{4}
\subsection{Sky localization performance}
\label{4_1}
\begin{figure}[t]
\begin{minipage}{18pc}
\includegraphics[width=18pc]{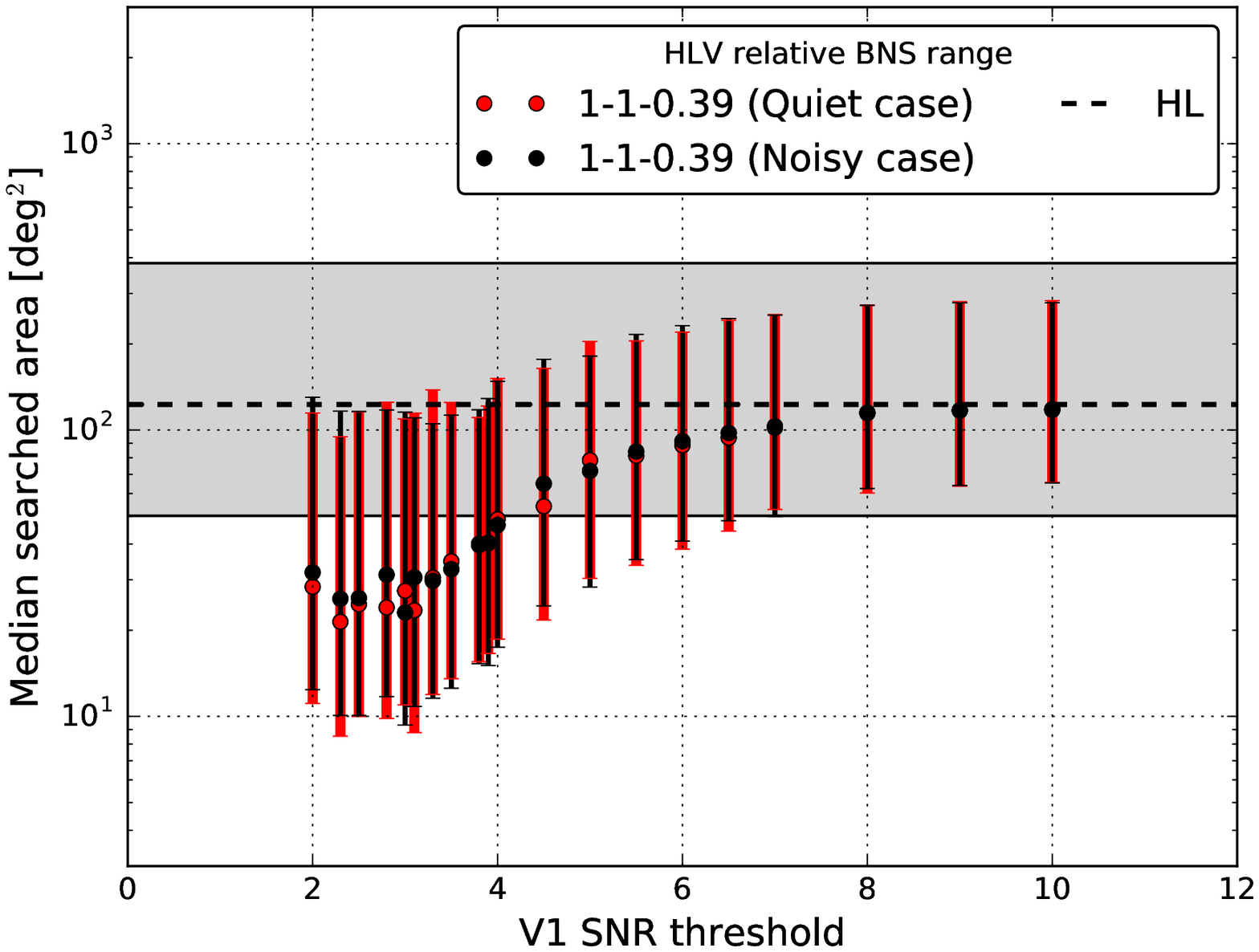}
\end{minipage}\hspace{-1.5pc}
\begin{minipage}{18pc}
\includegraphics[width=18pc]{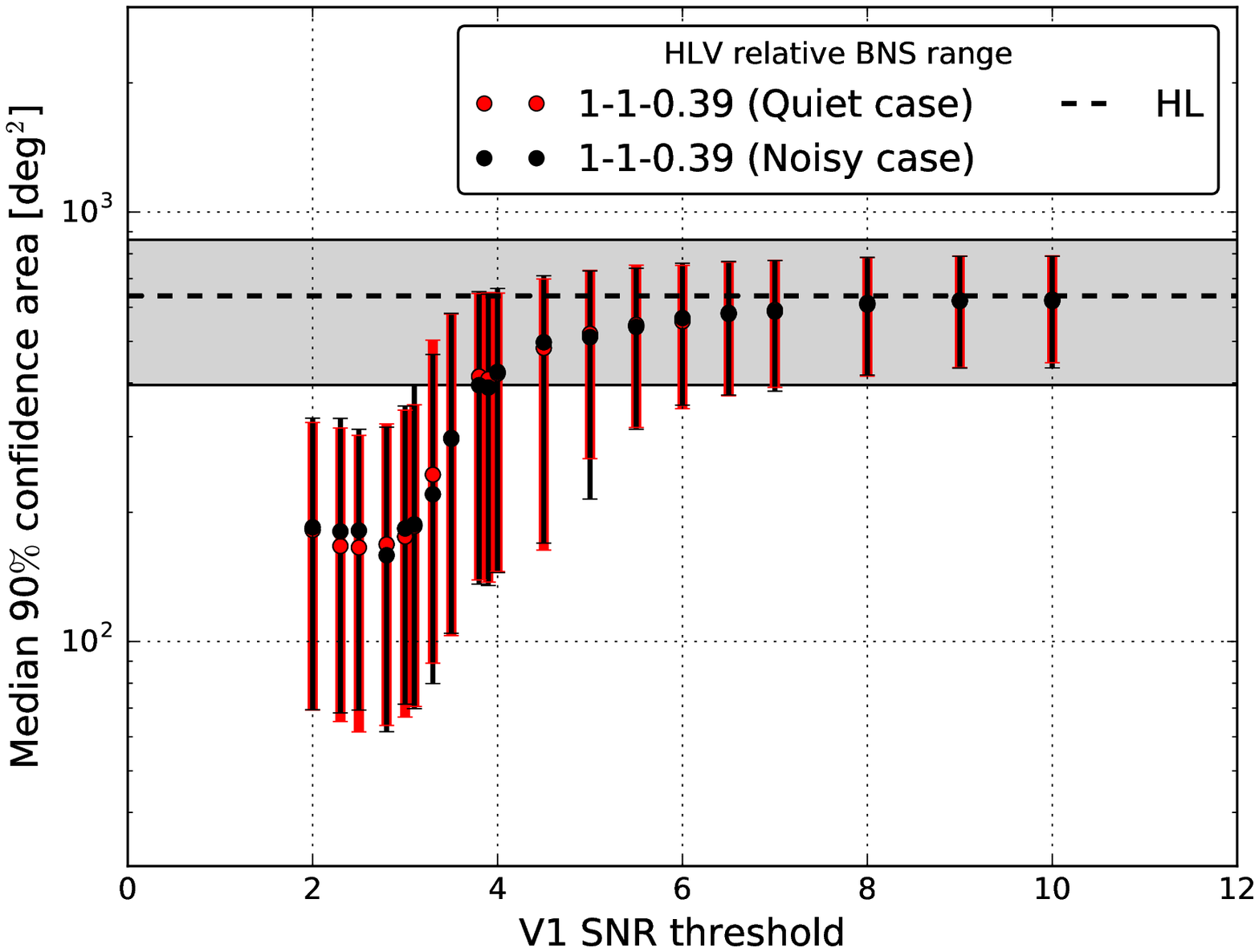}
\end{minipage}
\caption{\label{fig4_1_2}The sky localization performance of the heterogeneous network search by three detectors.
The dashed black line shows the performance using the two LIGO detectors only.
The red colored dots are for the case when the detector is quiet, while the black ones correspond to the noisy condition.
The performances in the two cases are similar.}
\end{figure}

Using the above settings, we generate sky maps of 248 events and investigate the localization performance by collecting the median values of searched area and 90$\%$ confidence area.
This calculation is repeated with different SNR thresholds in Virgo.
First, we assumed that the two LIGO detectors have the same range while the Virgo detector has 39$\%$ of the range of the LIGO detectors to roughly mimic the O2 sensitivity.
The calculated performance is shown in Figure \ref{fig4_1_2}.
The dots show the median of the localization areas over the set of injections and the uncertainties report the interquartile range.
In order to check that the results were not overly sensitive to a particular realization of the random numbers used in the simulation, the procedure was repeated twice, and since the results were consistent, the figure reports the average (quadratic average) of the medians (uncertainties) obtained in the two trials.
The relative detector sensitivities are written down as 1-1-$x$, with $x$ the ratio of the Virgo sensitivity compared to the two LIGO detectors.
The plots show that the optimal SNR threshold in Virgo is around 3. At this threshold, the localization is improved by about a factor of 4.
In this configuration, about 51$\%$ of the 248 events are reconstructed as HL${\rm V}_{\rm i}$ triggers, 36$\%$ as HL${\rm V}_{\rm n}$ triggers, and 13$\%$ are HL triggers.
The percentage of these three triggers changes depending on the Virgo threshold
(these percentages depend on the Virgo sensitivity and threshold).
When running MBTA on Virgo data with a SNR threshold ranging from 5 to 10, we obtained consistent median values for the searched area and 90$\%$ confidence area compared to the localization performance obtained with artificial Virgo triggers at those threshold as shown in Figure \ref{fig4_5}.
Figure \ref{fig4_1_2} also includes the localization performance when the lower sensitivity detector generates louder background triggers compared to a stable case.
In this noisy case investigation, the SNR distribution and the FAP in Virgo are derived from the black curves instead of the red curves in Figure \ref{fig3_2_1}.
This has no impact on the localization performance and in the following we work only in the quiet case.

\subsection{Dependence on the sensitivity of the third detector}
\label{4_1}
\begin{figure}[t]
\begin{minipage}{18pc}
\includegraphics[width=18pc]{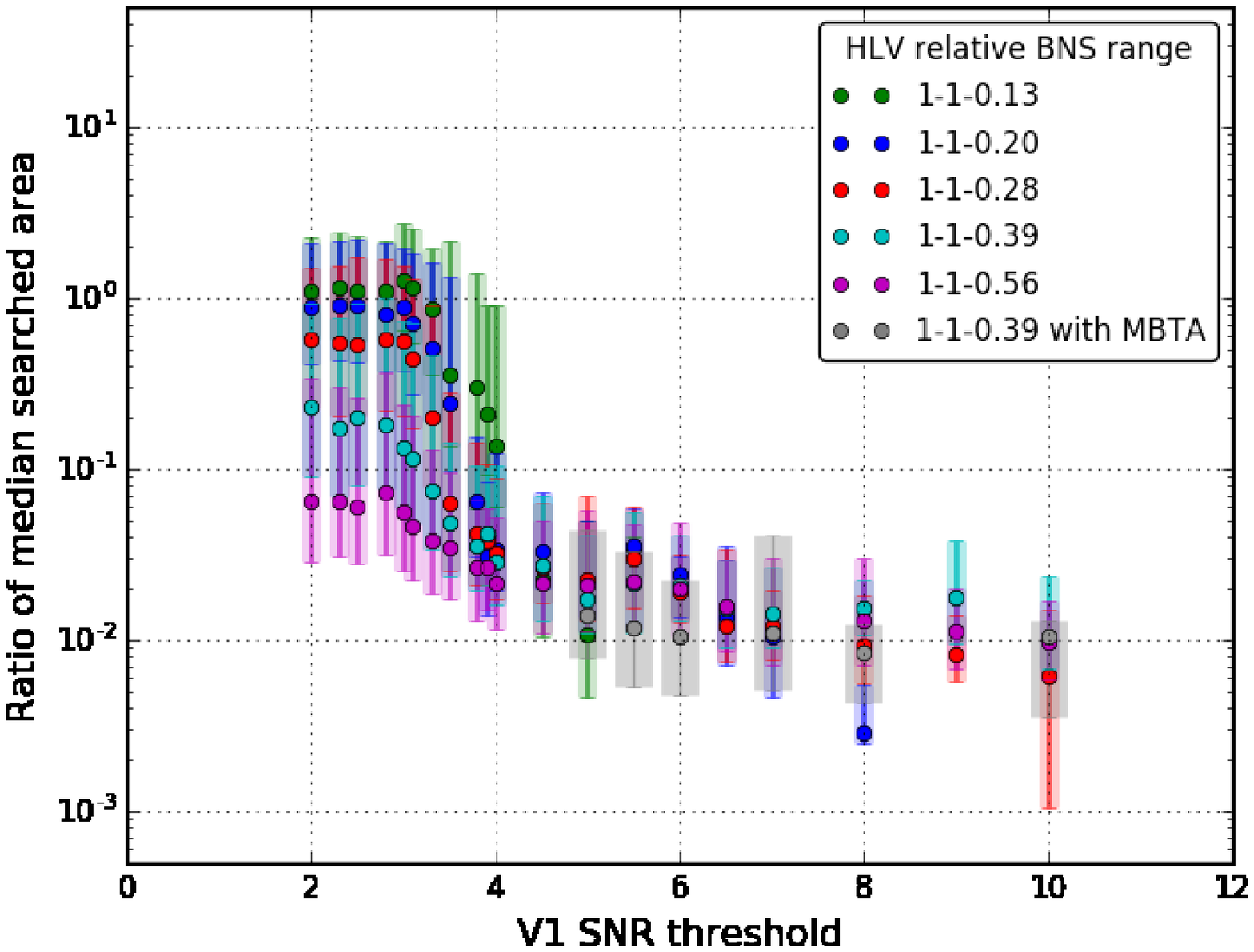}
\end{minipage}\hspace{-1.5pc}%
\begin{minipage}{18pc}
\includegraphics[width=18pc]{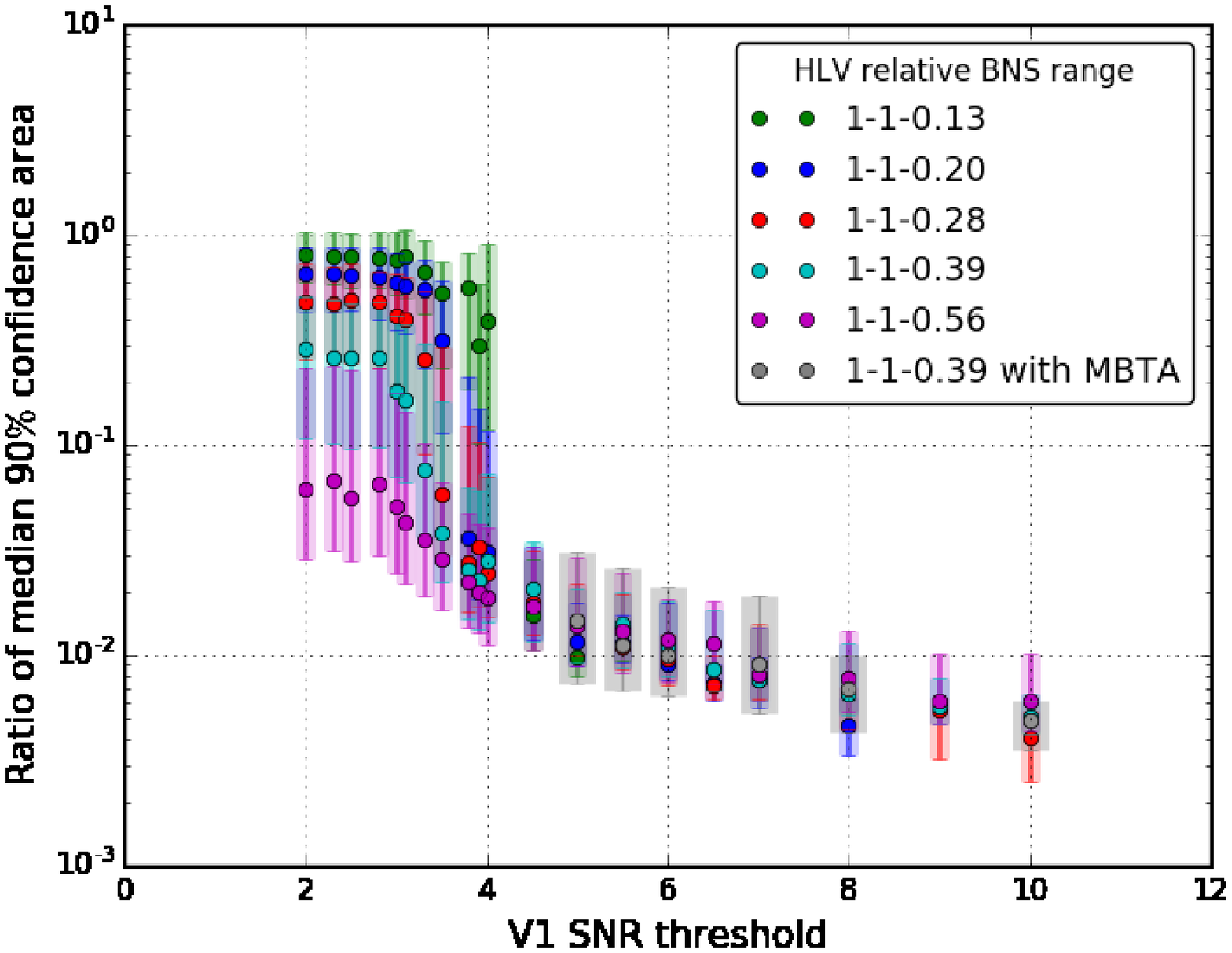}
\end{minipage}
\caption{\label{fig4_5}Ratio of the median searched area ($left$) and 90$\%$ confidence area ($right$) of HLV triggers to that of the same triggers treated as HL coincidences.
As the sensitivity improves, HL${\rm V}_{\rm n}$ triggers become less likely whereas HL${\rm V}_{\rm i}$ triggers become more likely and benefit the localization performance. 
}
\end{figure}

\begin{figure}[h]
\begin{minipage}{18pc}
\includegraphics[width=18pc]{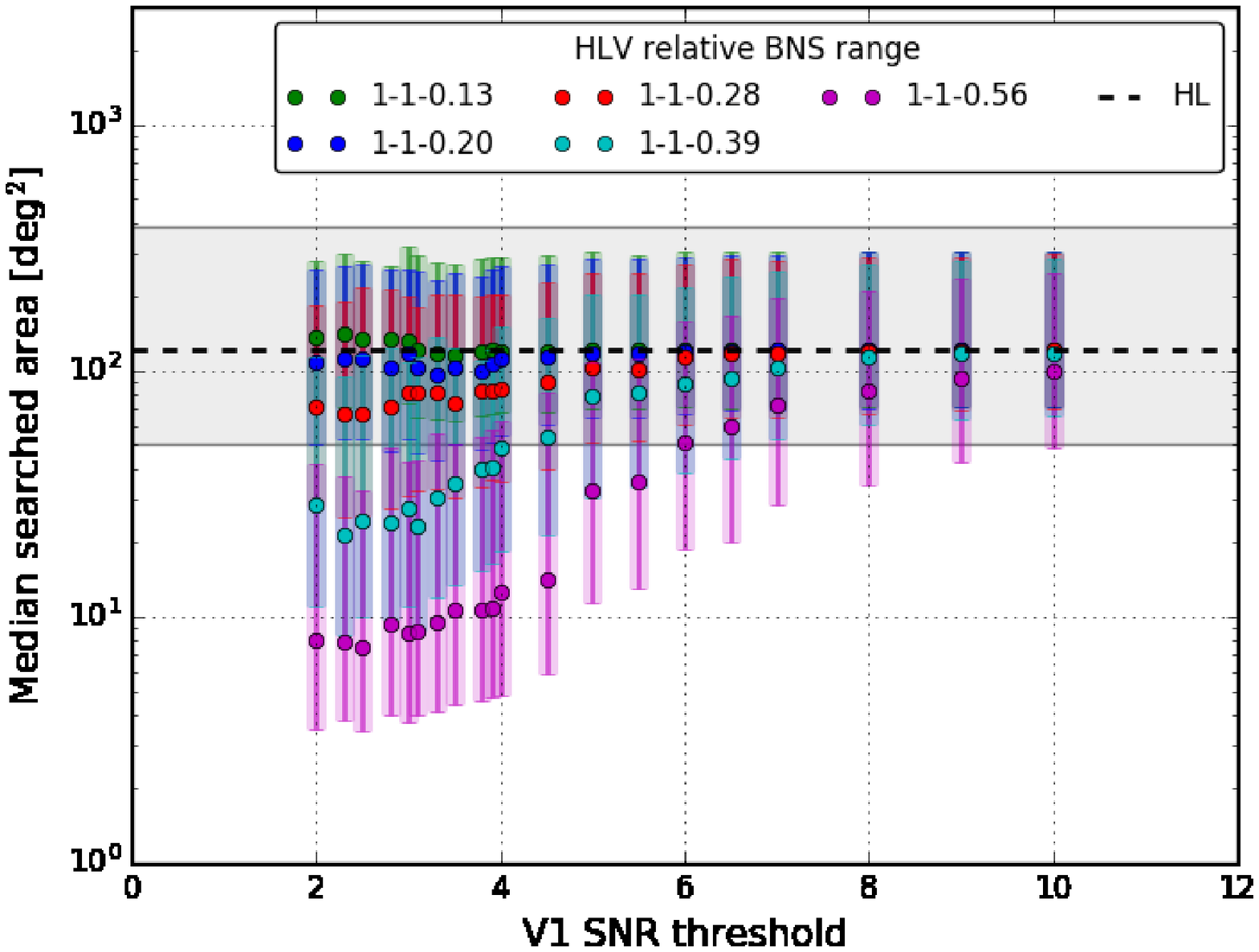}
\end{minipage}\hspace{-1.5pc}
\begin{minipage}{18pc}
\includegraphics[width=18pc]{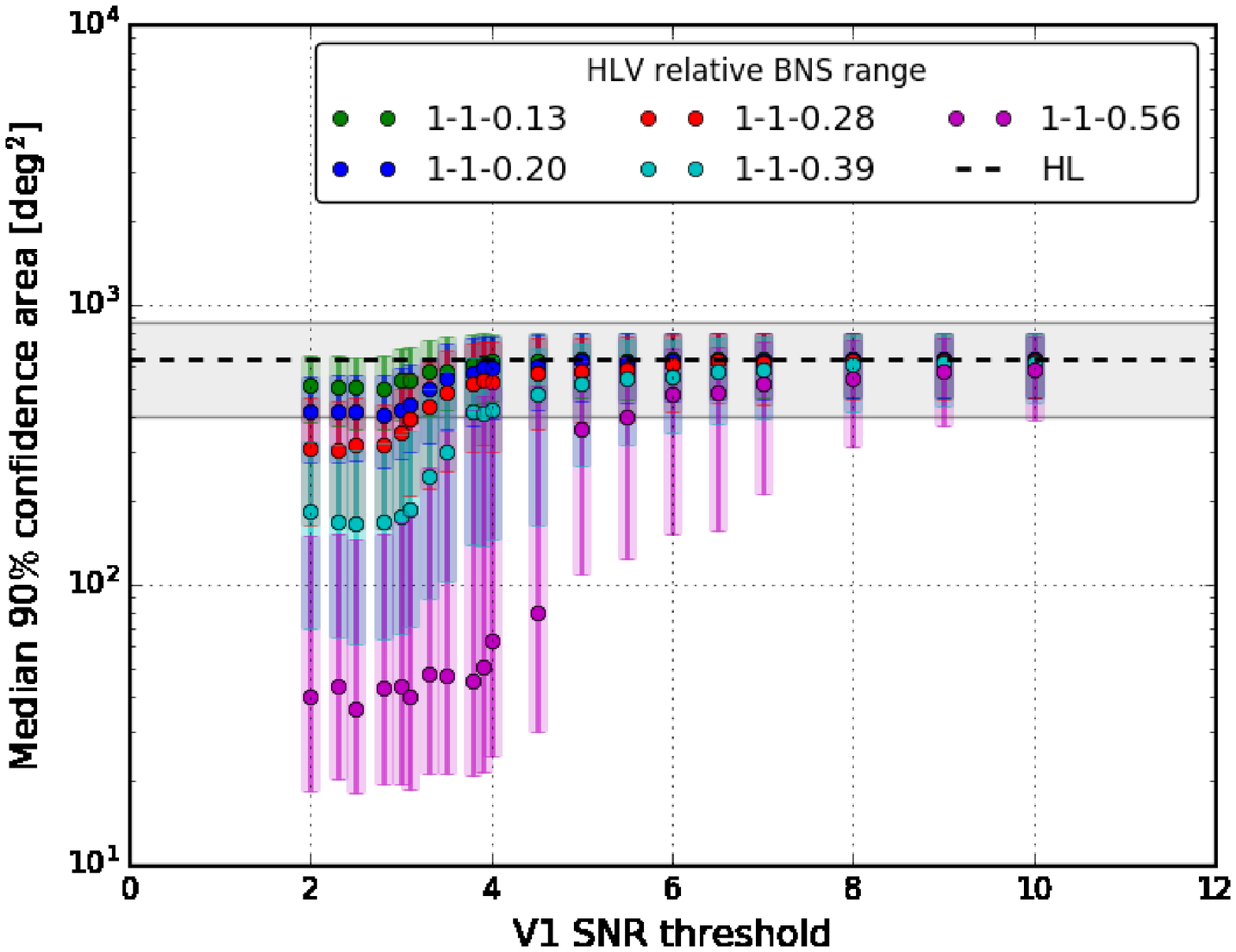}
\end{minipage}
\caption{\label{fig4_4}The median searched area ($left$) and 90$\%$ confidence area ($right$) are shown as a function of the SNR threshold used in Virgo.
Expected sky localization performance with the hierarchical search when the sensitivity of the Virgo detector is varied.
The colors show the network configuration.}
\end{figure}

We calculate the localization performance for various BNS ranges of the Virgo detector.
This is realized by scaling its expected SNR using (\ref{eq3_5_1}) and its noise curve.
The other settings are as described in section \ref{3_3}.
The expected performances in terms of searched area and 90$\%$ confidence area are shown in Figure \ref{fig4_4}.
The hierarchical search with three detectors will improve the precision of the localization regardless of the value of the BNS range.
On the other hand, if the BNS range for the lower sensitivity detector is too low, the accuracy of the localization will be slightly degraded since it will be more likely to get triple coincidences with Virgo noise triggers.
However, this effect will not be so large; noisy Virgo triggers will mainly be found when the sensitivity is much lower than the sensitivities of the LIGO detectors.
In this situation, the reconstructed sky maps will be similar to the ones obtained with the LIGO detectors only.
Figure \ref{fig4_5} shows the improvement in the performance for those events that become triple coincident events when Virgo is added to the network.
Based on Figures \ref{fig4_5} and \ref{fig4_4}, the lower sensitivity detector begins to improve the localization performance as soon as its sensitivity is 20$\%$ of the more sensitive detectors.
Then the optimal SNR threshold lies in the range from 3 to 3.5.
We find that as the sensitivity of Virgo approaches one third of the LIGO one, the accuracy and precision of the localization can be improved by about a factor of 3.

Concerning the sky maps, the reconstructed region by HLV coincident triggers will be a fraction of the area reconstructed by the two LIGO detectors only.
If a coincident trigger is built from two LIGO signals and a noise Virgo trigger, the area pointed by the HLV network starts to shift from the source position, mostly due to the error on the detection time in Virgo.
However, if the range of Virgo detector is much lower than the LIGO ones, the reconstructed area remains similar to the ring shape region reconstructed by the two LIGO detectors only even when the Virgo trigger is due to noise.
This hierarchical search will find ring-shaped sky maps when the sensitivity of the third detector is much lower than the higher sensitivity ones.
As the sensitivity improves, the sky maps progressively turn into point-like regions inside the area by the two LIGO detectors.

\section{Conclusion}
Using MBTA and Bayestar, we show the expected fast localization performance for GWs from compact binary coalescence when a hierarchical search is implemented into a GW-EM follow-up pipeline.
We confirm that the hierarchical search will improve both the localization accuracy and precision compared to those achieved by a double coincidence search with the two LIGO detectors alone.
The hierarchical network effectively improves the localization accuracy and precision when threshold SNR for the lower sensitivity detector is set to around 3.5 provided that the BNS range of that the detector is greater than 20$\%$ of more sensitive detectors.
Consequently, this hierarchical search will be most useful when adding new, less sensitive detectors to the network, as they are undergoing commissioning.

\section*{Acknowledgments}
We thank Leo Singer for providing help with Bayestar. We thank the MBTA team at LAPP and Universit\`a degli Studi 
di Urbino for useful discussions. TA and YF have received support from Labex Enigmass for this work.
This work was also supported by JSPS Core-to-Core Program A. Advanced Research Networks, and JSPS Grant-in-Aid for JSPS Fellow 17J03639.




\bibliographystyle{elsarticle-num-names}
\bibliography{bibliography}





\end{document}